# Effect of strain path on deformation texture of superconducting niobium sheet

A. Zamiri

Mechanical Engineering Department, Michigan State University, East Lansing, MI 48824-1226

**Abstract**

The texture of high purity superconducting niobium sheets plays an important role in the physical and mechanical properties of high purity niobium sheet that are important for manufacturing of superconducting accelerator cavities. In a particular batch of sheet metal, orientation imaging microscopy showed an inhomogeneous texture from the surface to the mid-thickness of the sheet consisting a $\gamma$ fiber, {111}<uvw>), cube fiber, {100}<uvw>), and also some components on the $\alpha$ fiber, {hkl}<110>. With uniaxial deformation, peaks on the {uvw}<111> $\gamma$ fiber evolve differently depending on the in-plane direction of deformation, and the position in the sample (surface vs. center). Applying a different strain path such as balanced biaxial bulging, leads to development of rotated Goss, {110}<110> components in the texture of the deformed niobium. These results show that the texture of niobium is very sensitive to the deformation and strain path.

## 1. Introduction

High purity niobium sheets are deformed to large strains during the manufacture of radio-frequency superconducting accelerator cavities that enable acceleration of a charged particle beam. There are many factors that affect the performance of these cavities. The microstructure of niobium has a significant effect on the etching processes required to obtain a clean and smooth surface that is essential for high performance of accelerators [1]. Due to the need to develop larger fields for advanced accelerators, investigation of the microstructure and the properties of niobium to make it more formable for making more complex geometries without negatively affecting the required surface physical (functional) properties has become increasingly important.

The mechanical properties of high purity niobium have been studied in the past. The anisotropic behavior of high purity niobium is very sensitive to the amount of the plastic deformation [2]. Therefore, to appropriately predict the complex plastic behavior of high purity niobium, an evolving yield function which accounts for the microstructural evolution should be used [3,4]. The surface texture and microstructure has been observed to have an important role in controlling the deformation induced surface roughness and some physical properties such as field emission and phonon transport in high purity niobium [5].

One of the important characteristics of the microstructure that governs many properties of materials is its crystallographic texture (distribution of grain orientations) in the microstructure. The anisotropic mechanical and physical behavior of a single crystal makes the crystallographic texture an important parameter to enable control of mechanical and physical properties. BCC materials tend to develop fiber- type textures during deformation and annealing [6-8]. The most important fibers in BCC materials are RD or α-fiber consisting of orientations with a common <110> direction parallel to RD (rolling direction) and the ND γ-fiber which consists of {111}<hkl> orientations, where {111} planes lie in the rolling plane. Additional fibers that contain some important orientations are the cube fiber or {100}<hkl>, the ζ-fiber containing {110}<uvw> and the ε-fiber which contains the {554}<225> orientation that is important in steels. Many investigations of BCC materials show that the intensity of different orientations along these fibers controls the formability of the material [7-10]. The ratio of {111}/{100} intensities in the ND should be high to obtain good deep drawability of BCC sheet metals [9,10]. Besides formability, texture is an important parameter in controlling some surface physical properties such as electron field emission arising from surface irregularities arising from differential etching rates of different grain orientations from the inner surface of cavities [1]. Consequently, variations in texture with depth affect forming properties, and surface texture affects etching processes that ultimately govern the performance of the cavity.

Due to the lack of knowledge about the texture and texture development in high purity niobium sheets, this study was undertaken to quantify the texture in highly pure niobium sheets. The texture of the undeformed (initial or as-received sheet), uniaxially deformed tensile specimens, and balanced biaxial bulged niobium were measured using orientation imaging microscopy (OIM) to identify through-thickness texture gradients and also the deformation texture of the sheet.

## 2. Experimental measurements and methods

The chemical composition of superconducting niobium used in this study is given in Table 1. The material had been rolled to 2 mm thick and recrystallized by the supplier (Tokyo Denkai). Tensile specimens were cut using electron discharge machining (EDM) with the tensile axis in five different directions with respect to the rolling direction Tensile experiments were conducted using specimens based upon ASTM **E 517**, using a strain rate of 0.005 s$^{-1}$. Following deformation, selected specimens were sectioned perpendicular to the tensile direction, mechanically polished, and electropolished for EBSP measurements on the cross-section. To investigate the effect of a different strain path on deformation texture of high purity niobium, a balanced biaxial bulging test of niobium sheets were also conducted at 0.005 s$^{-1}$ strain rate and specimens were cut for texture investigation at different orientations as indicated in Figure 1.

Texture was measured using OIM scans across the full thickness of the specimen on a CAMSCAN 44FE microscope using a hkl Technologies EBSP mapping system. The scans sampled about 600 grains. Post processing was accomplished using TexSEM analysis software version 3.0, using nearest neighbor correlation methods to clean up the data set (to change pixels with low confidence in correct indexing with a higher confidence neighboring pixel orientation). Orientation distribution functions (ODFs) were calculated from the discrete data without imposing any (orthotropic) sample symmetry assumptions.

Table I  Chemical composition of superconducting niobium

| Element | Ta | W | Ti | Fe | Si | Mo | Ni | Zr | Hf | O | N | C | H |
|---|---|---|---|---|---|---|---|---|---|---|---|---|---|
| Composition (wt%) | 0.024 | <0.002 | Each <0.001 | | | | | | | | | | <0.0002 |

## 3. Results

*3.1. Texture analysis after uniaxial tension*

OIM measurements on the as-received material showed a texture gradient from the surface to mid-thickness of the high purity niobium sheet. This texture gradient is most apparent in color inverse pole figure orientation maps aligned with the normal direction, but they are also visible in the crystal direction maps in figure 2, which also illustrates the grain morphology in the undeformed specimen. Due to the greatest variation of the r-value at 45°, 67.5°, and 90°, compared to the rolling direction [2], the texture of these uniaxially-deformed specimens were also measured and compared with the undeformed sample. To assess the texture gradients, three equal area layers denoted as surface layer, subsurface layer, and the mid-thickness layer are compared.

The most important information about the texture components of BCC materials can be viewed on the 45° sections of ODFs. Figure 3 shows a 45° section of the ODF and important fibers and orientations in that section. There are three fibers of particular interest on this section: 1) the α-fiber, on the left vertical axis ($\varphi_1 = 0$) consisting of orientations with a common <110> direction parallel to the rolling direction ({uvw}||<110>); 2) the γ-fiber, represented by a horizontal line running along $\Phi = 54.7°$ consisting of the <hkl>||{111}orientations, where the {111} planes lie in the rolling plane and are therefore perpendicular to the normal direction; and 3) the ε-fiber, represented by the right vertical axis ($\varphi_1 = 90$) consisting of orientations that lie between {001}<110> to {110}<001> and containing important orientations near {554}<224>. Additional important fibers in BCC materials are also located on this section of the ODF. The first is the cube fiber, which corresponds with the upper horizontal axis of all sections of the ODF, and contains the orientations for which {001} is parallel to the rolling plane; it can therefore be described as {001}<uvw>. The second is the {110}<uvw> fiber, which has two points on the fiber at the bottom corners of the $\varphi_2 = 45°$ section where {110} is parallel to the rolling plane. Mention the eta fiber too?

After deformation, the ODF plots were qualitatively similar, but there were significant differences in the orientations of peaks along the fibers as illustrated in figures 4 to 14. Figures 4,6,8 provide $\Delta\varphi_2 = 15°$ sections of the in each of the three layers in the as-received specimens (the texture varies smoothly between the sections shown). Skeleton plots of important fibers illustrate how deformation in different directions altered the intensities more effectively than ODF plots. Details of the texture investigation in different layers and also the effect of plastic deformation on texture evolution are discussed layer by layer in the following sections. In all skeleton plots, the as-received coordinate system is the basis for presenting changes that arise from deformation.

*3.1.1. Mid-thickness layer texture*

Figure 4 shows ODF sections for mid-thickness layers of as-received niobium. As is clear from the 45° section of ODF, there is a strong γ-fiber present in the mid-thickness layer texture of the as-received niobium while there is almost no intensity for the α-fiber in the texture. The skeleton diagrams for α and γ-fibers of as-received and deformed sheet are shown in figure 5, indicating strong peaks on the γ-fibers in all as-received and deformed niobium sheets. The as-received material and the uniaxially stretched specimen in 67.5°, show a peak of intensity near the {111}<011> component on the γ-fibers. The uniaxially deformed specimen in 45° and 90° show a very intense peak near the {111}<112> component. The α-fiber is very weak in all specimen orientations (figure 5b). There is no evidence of {100} component, cube orientations, {110} components, or Goss orientations.

The above observations show that the direction of plastic deformation has a significant effect on the final mid-thickness texture of the deformed niobium sheet.

*3.1.2. Subsurface layer texture*

The subsurface layer ODF of the as-received niobium is shown in figure 6, indicating a modest γ-fiber with spreading onto the ε fiber. The α-fiber is very weak and there is a moderate intensity for {100} or cube orientations evident along the top edge of all sections of the ODF. The skeleton diagrams of as-received niobium (figure 7a) shows an almost continuous γ-fiber containing components between {111}<011> and {111}<112>. In the specimens deformed in the 45° and 90° directions, a γ-fiber peak has an orientation around {111}<112>, whereas the peak in the γ-fiber in the 67.5° direction is around {111}<011>. The α-fiber is weak in all deformed and as-received sheets (figure 7b), but there is some intensity along the α-fiber in the 90° direction between {001}<110> and {113}<110>, between {112}<110> and (111)<110> for the 67.5° direction, and at {223}<110> for the 45° specimen.

Compared to the mid-thickness layer, the γ-fibers of both deformed and as-received sheets are weaker in the subsurface layer while the α-fibers are stronger, and there is moderate intensity for {100} or cube orientations.

The above observations show that with plastic deformation the intensity of some specific components on the γ-fiber of subsurface layer of niobium increase significantly which depends on the direction of deformation. The α-fiber also becomes stronger with plastic deformation in subsurface layer of niobium.

*3.1.3. Surface texture*

The surface layer texture of superconducting niobium affects important physical properties of niobium sheets.

The ODF (figure 8) of the surface layer in the as-received sheet shows a rather weak γ-fiber in the surface layer with only two peaks on it. The $\varphi_2 = 0°$ section of the ODF shows {100} cube orientations in the surface layer texture that are stronger than in the subsurface layer. The strong intensity for near-cube orientations {015}<100> is also clear in the η fiber ($\varphi_1 = \varphi_2 = 0$) shown in figure 9. Figure 10 shows the skeleton diagrams for the surface layer texture of deformed and as-received niobium sheets. As is shown in figure 10a, a rather weak γ-fiber with a peak of intensity around the {111}<011> component is evident in the as-received sheet. Similar to the peak on the η fiber, the α-fiber also has a peak close to the {115}<110> orientations.

Uniaxially deformed specimens in 45° and 90° directions show strong peak type γ-fibers with a peak of intensity around {111}<112> components. The intensity and the shape of the γ-fibers in these specimens are almost the same as in the subsurface textures. The α-fiber in deformed specimen in the 45° direction shows a rather strong peak of intensity around {223}<110> component.

The shape of the γ-fiber in the surface layer of the deformed sheet in the 67.5° direction is almost the same as in the subsurface layer but with less intensity for {111}<110> component. The α-fiber is rather weak in this specimen (figure 10b).

The surface layer texture of the as-received material consists of a rather weak γ-fiber and some strong peaks on the α-fiber, which evolve significantly with the plastic deformation, and differently according to the deformation direction. With deformation, the intensity of most of the components along α-fiber and γ-fiber decrease to strengthen some specific components along γ-fiber.

*3.2. Texture of bulged niobium*

To examine how deformation texture evolution in high purity niobium depends on strain path, the through thickness texture of balanced biaxial bulged niobium was also investigated. An inhomogeneous texture was observed in all directions in the bulge specimens. Figure 11 for example, shows an inhomogeneous through-thickness texture in 0° direction (compared to the rolling direction) in the bulged niobium sheet.

*3.2.1. Surface texture*

As discussed before, the surface texture of as-received niobium includes a strong {115}<011> component on α-fiber, strong {100} components, and a relatively weak γ-fiber with a peak of intensity around {111}<011> orientations. Figures 12a and b show the skeleton diagrams of the surface of bulged niobium sheet and the as-received undeformed niobium. The γ-fibers of the bulged niobium samples also show a peaked fiber with a maximum intensity around the {111}<011> orientation as undeformed material. As is clear from figure 12a, biaxial deformation caused an increase in the left {111}<011> component in all directions, but it is the highest in the $0^o$ direction and weakest in the 45° direction. The two {111}<011> components have equal strength for the 90° orientation, which preserved the original peak on the γ-fiber, in contrast to tensile deformation, where {111}<112> components became strong. The α-fibers also show peaked intensities near the intersection with the γ-fiber between near $45^-55^o$, similar to the tensile experiments, but a peak also developed at 90°. The initial {115}<011> intensity was lost while orientations close to {111}<011> and rotated Goss, {011}<011>, were strengthened in the surface texture. Detailed study of the ODF did not show any evidence of {100} or cube components in the surface texture of bulged niobium sheet.

*3.2.2. Subsurface texture*

The γ-fiber of subsurface texture is also a peak type fiber with intensity around {111}<011> orientations (figure 13a). The γ-fibers in all 5 directions are weaker than the as-received material and like the surface texture the γ-fiber is strongest in the $0^o$ direction. The α-fibers in all 5 directions are also peak type and it is stronger in $0^o$ direction (figure 13b). Comparing these fibers with as-received material, one can see that with balanced biaxial bulging of niobium the rather strong orientations on the γ-fiber are consumed in favor of new Goss orientations, {110}<110>, on the α-fibers.

*3.2.3. Mid-thickness texture*

The γ-fiber of the mid-thickness texture layer of bulged niobium is very weak compared to the as-received material (figure 14a). The γ-fiber of deformed material is peaked and as like other two texture layers is stronger in $0^o$ direction. The γ-fibers show a peak of intensity around {111}<011> orientations. The α-fibers are much stronger in the deformed material, especially in $0^o$ direction with the peak of intensity around $45^o$ and $90^o$ orientations (figure 14b). Therefore, with bulging of niobium the intensity of the orientations along the γ-fiber decrease in favor of an increase on orientations around $45^o$ and $90^o$ (rotated Goss orientations, {110}<110>) orientations along α-fiber.

**4. Discussion**

As-received niobium, as discussed above, has a strong cube component, {100}<001>, on the surface. The texture of the surface layer of the sheet metals develops differently from the interior due to closer proximity to the enhanced shear deformation due to friction with the rolling mill, or due to small reductions that concentrate deformation in the surface layer. Usually, the presence of the intense cube {100}<001> orientations on the surface texture of BCC materials is related to the friction between tools and the sheet during deformation [11-12,17,18].

It has been observed that usually the recrystallization or deformation texture of sheet metals depends on the initial texture prior to recrystallization or deformation [7,9,11]. To obtain a desirable final texture in material it is necessary to control the texture development from initial state of deformation and processing [13-16]. The above investigation shows that the deformation texture differences through the thickness layers of niobium sheets completely depends on the prior texture in recrystallized material. The mid-thickness layer, which has a strong texture components in the γ-fiber, leads to a texture consisting of strong components in the γ-fiber after deformation, while the near surface or surface textures, which have weaker γ-fiber, convert to a weaker γ-fiber (compared to mid-thickness texture) after deformation. Strong γ-fiber seems to prevent the development of rotated cube orientations during deformation.

The inhomogeneity of the texture through the thickness of the recrystallized sheets is related to the friction between the tools and the sheet during deformation and shear texture close to the surface of the sheet, which develops during the hot and cold rolling [17,18]. Purity is an important parameter in the superconducting niobium sheets, and this make the primary forming of niobium more complicated. The other issue that introduces inhomogeneous through thickness texture in sheets is using an insufficient deformation zone depth during cold rolling of the sheets, and this happens when the plate prior to cold rolling is very thick or when the rolling reduction in any pass is not enough.

Many investigations on BCC metals have shown that there is a direct relationship between the deep drawability and planar anisotropy and texture of the metal [7-10]. Usually, a weak α-fiber and a strong continuous γ-fiber in the texture of a material are desirable for the formability of sheet. As a general rule it has been documented that deep drawability is directly related to the ratio of {111}/{100} components of the texture. The mid-thickness texture of both as-received and deformed materials consists of strong {111} orientations and there is no intensity for {100} orientations. The γ-fiber of as-received niobium is more continuous while after deformation, strong peaks develop along the gamma fiber. From mid-thickness to the surface, both the intensity and continuity of the γ-fiber decreases in the as-received niobium while strong peaks develop in the deformed materials (figures 5,7,10,12-14). These observations suggest that one possible mechanism to explain the complex anisotropic deformation behaviors of niobium, reported in [2], could be the development of a peaked γ-fiber which only contains a very intense component around a particular orientation in the texture of the deformed materials.

Planar anisotropy of BCC materials is related to the intensity of different components in the texture of material. Ideally, a texture consists of a pure and continuous γ-fiber with a weak α-fiber to introduce a zero planar anisotropy to the material. A higher amount of planar anisotropy in materials leads to a non-symmetric cross section in the final deep drawn parts with non-uniform thickness. In superconducting accelerator cavities made from niobium, geometry is an important parameter in the final parts. Thickness variations where welding occurs may lead to the need to vary welding parameters to prevent welding defects. Therefore, keeping the planar anisotropy as small as possible is important in order to allow robust processing and fabricating conditions, which can be obtained by controlling the texture components in the recrystallized material.

Planar anisotropy can be classified in two types, V-type planar anisotropy and specific planar anisotropy [19]. V-type planar anisotropy happens when the r-values in 0 and 90°, compared to the rolling direction are maximums. Specific planar anisotropy happens when one of the r-values in 0 or 90° is smaller than the r-values between 0 and 90°. As was shown in [2], superconducting niobium sheets show a specific planar anisotropy. The appearance of specific planar anisotropy in as-received high purity niobium sheet may be related to strong peaks on the γ-fiber and also the presence of {100} and {115}<110> components especially in the subsurface and surface layers. As reported in [2,4], at small plastic strain the r-values of the niobium show a maximum at 67.5° and minimums at 45°and 90° but with plastic deformation the r-values become maximums at 45°and 90° and minimum at 67.5° directions. In the as-received materials the γ-fibers especially in the subsurface and surface layers show a peak of intensity around {111}<110> orientations and low intensity for {111}<112> components. But with an uniaxial plastic deformation in 45°and 90° direction, the intensity of the {111}<112> components increase significantly and all other components on the γ-fiber vanish while a uniaxial plastic deformation in 67.5° direction leads to a significant increase in the intensity of {111}<110> components. As is clear from figures 5, 7 and 10, after the plastic deformation the intensity of {111}<110> components in the γ-fiber of specimen stretched in 67.5° direction is less than the intensity of the {111}<112> component in the sheets stretched in 45°and 90° direction. This observation may suggest that the texture components is responsible for the r-value in 67.5° direction is {111}<110> while the r-values in 45°and 90° directions are affected by {111}<112> components.

## 5. Conclusion

As-received and deformed high purity niobium sheets show inhomogeneous through thickness textures which can be related to the deformation history of niobium sheets. The mid-thickness of as-received high purity niobium has a rather strong γ-fiber or {111}<uvw> fiber and a very weak α-fiber. Also, there is no {100} or cube orientation in the mid-thickness layers of niobium. From mid-thickness to the surface of niobium the intensity of the γ-fiber decreases while the intensity of {100} orientations increases especially in the layers close to the surface. There is a rather intense {115}<110> component in the surface layers of as-received high purity niobium.

The texture of high purity niobium is very sensitive to the plastic deformation and also to the type of strain path or the type of deformation. With plastic deformation, depending of the direction of deformation and the type of strain path, the intensity of the most of components on {111}<uvw> fiber decrease to increase the intensity of some particular components on the fiber. For example, in all three investigated thickness layers of niobium, the uniaxially stretching of niobium in 45° and 90° direction, compared to the rolling direction, decreases all the texture components on γ-fiber to develop strong {111}<112> orientations on the γ-fiber while stretching niobium 67.5° direction consume all the components to develop a strong component around {111}<110> orientations. Applying a different strain path such as balanced biaxial bulging leads to development of strong rotated Goss, {110}<110>, components on the α-fiber.

**Acknowledgment**

This work was supported by National Superconducting Cyclotron Laboratory (NSCL) at Michigan State University.

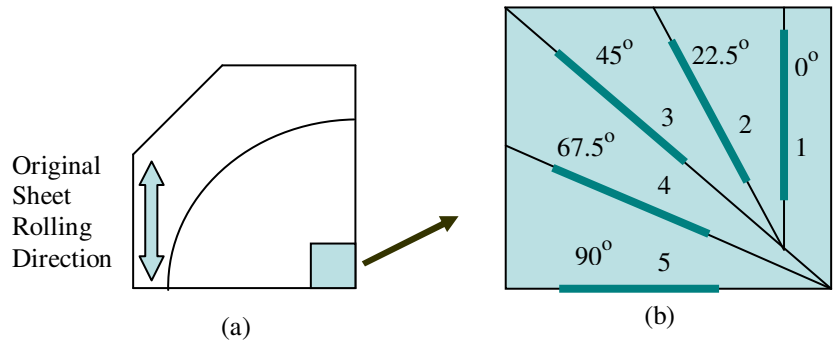

Fig. 1. a) One quarter of the bulged niobium part b) the locations and orientations of the specimens that were cut and analyzed using OIM.

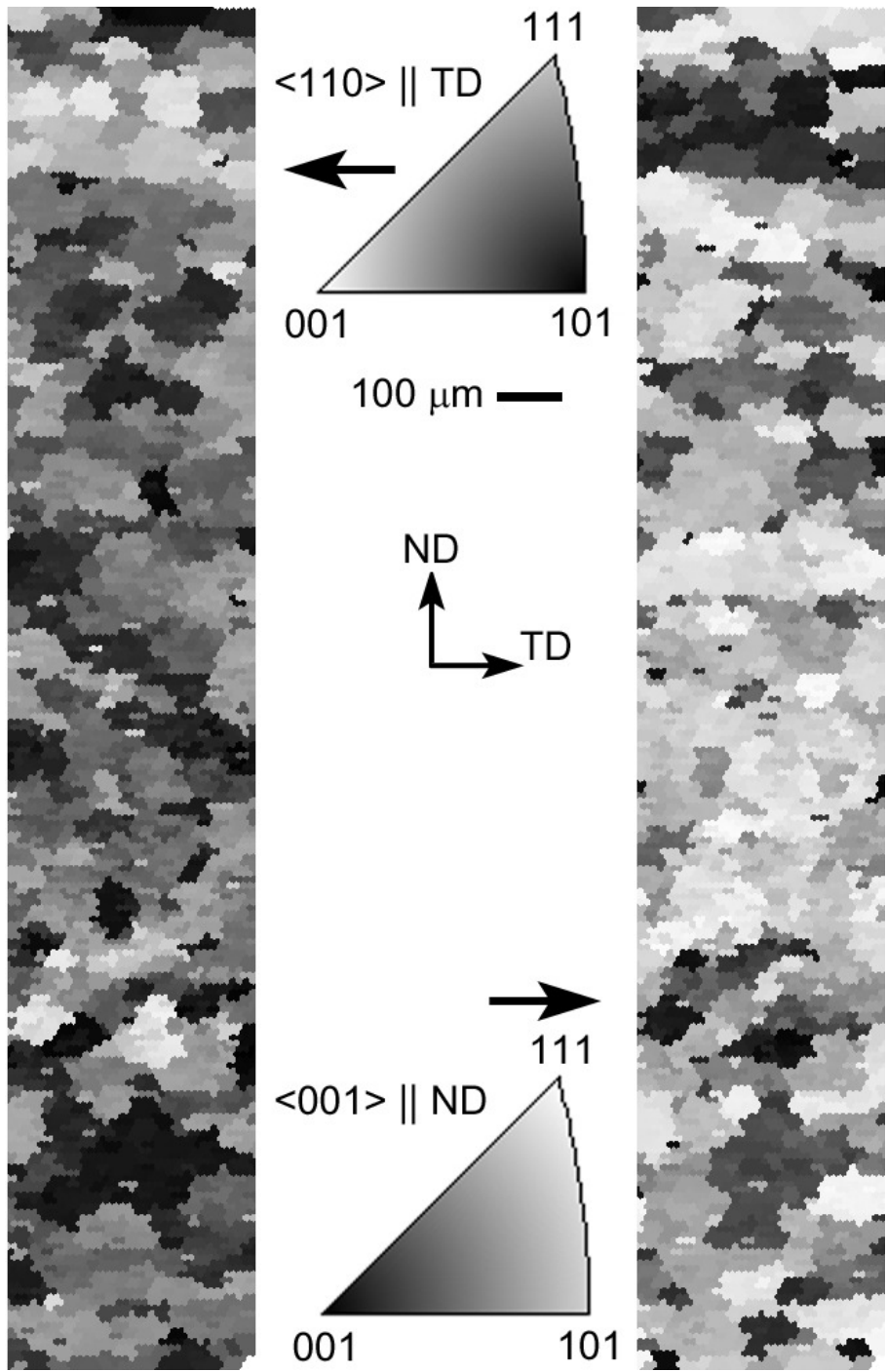

Fig. 2. Crystal direction maps from the same scan show the through thickness variation of texture in undeformed high purity niobium.

φ1

(001)[1$\bar{1}$0]  (001)[0$\bar{1}$0]  (001)[$\bar{1}\bar{1}$0]

← α-fiber   ε-fiber →

(113)[1$\bar{1}$0]
(112)[1$\bar{1}$0]
(223)[1$\bar{1}$0]

(111)[1$\bar{2}$1]   (111)[1$\bar{1}$0]   (111)[$\bar{1}\bar{1}$2]

(111)[1$\bar{1}$0]                                    (554)[$\bar{2}\bar{2}$5]

γ-fiber

(110)[1$\bar{1}$0]                                    (110)[001]

Fig. 3. Important fibers on φ$_2$=45° section of ODFs, and b) pole figures for undeformed and the fibers as they appear on a (100) pole figure. The pole figures are not here…

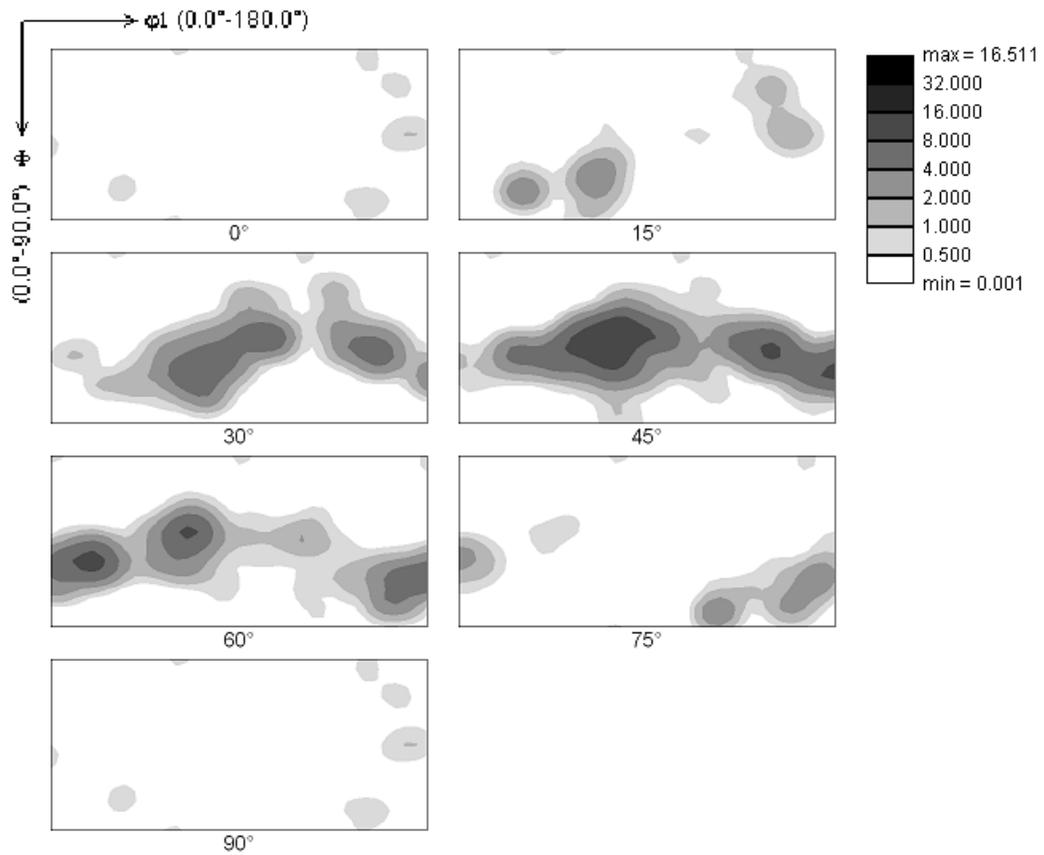

Fig.4. $\varphi_2$ sections of the ODF of the mid-thickness layer of undeformed niobium sheet.

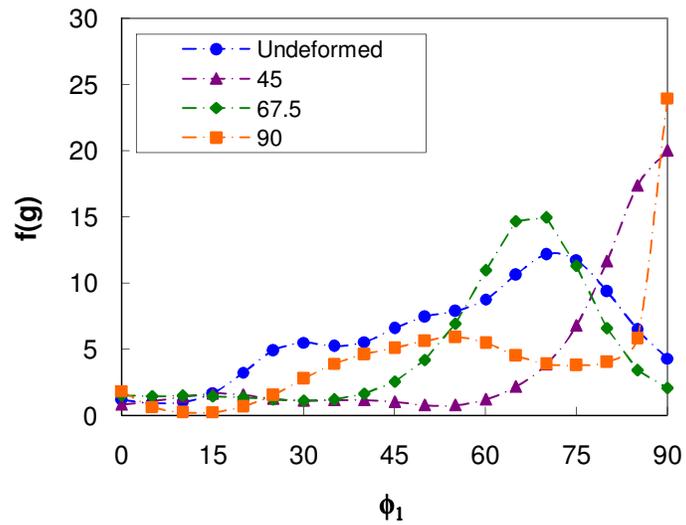

(a)

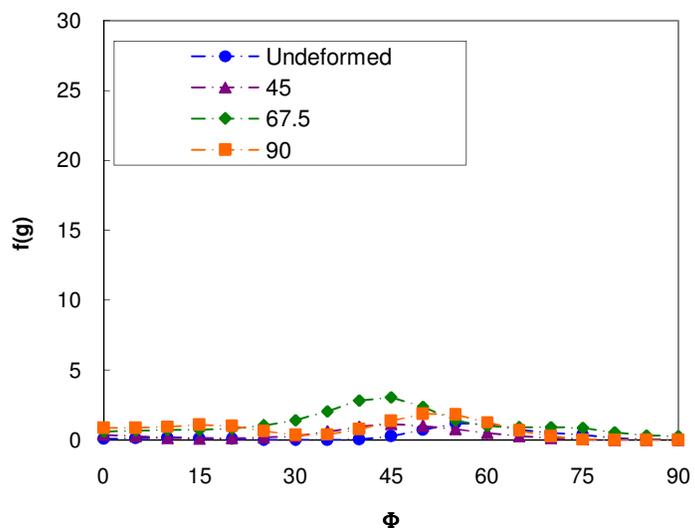

(b)

Fig. 5. The skeleton diagrams showing a) the $\gamma$ fiber and b) the $\alpha$ fiber of mid-thickness layer of the undeformed niobium and uniaxially deformed niobium in 45, 67.5, and 90° directions

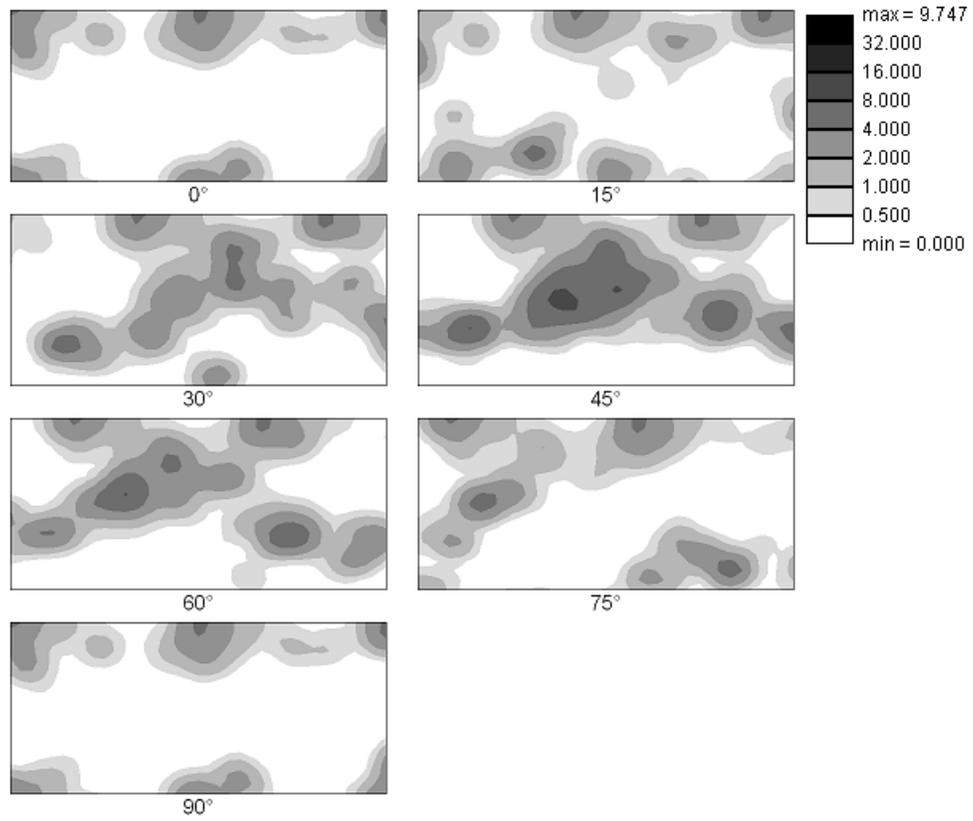

Fig. 6. $\varphi_2$ sections of the ODF of the subsurface layer of as-received niobium sheets

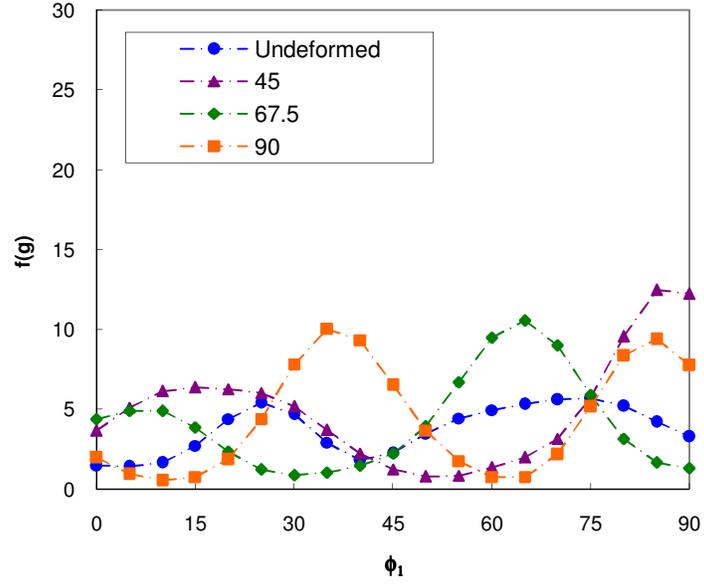

(a)

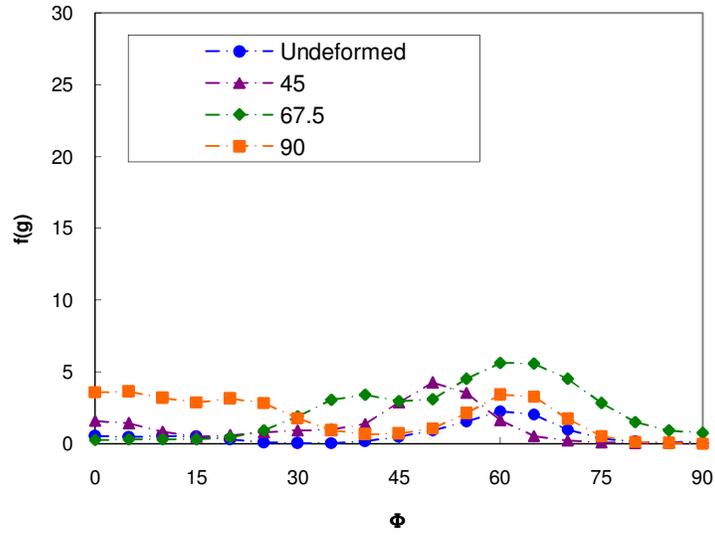

(b)

Fig. 7. The skeleton diagrams showing a) the $\gamma$ fiber and b) the $\alpha$ fiber of subsurface layer of the undeformed niobium and uniaxially deformed niobium in 45, 67.5, and 90° directions

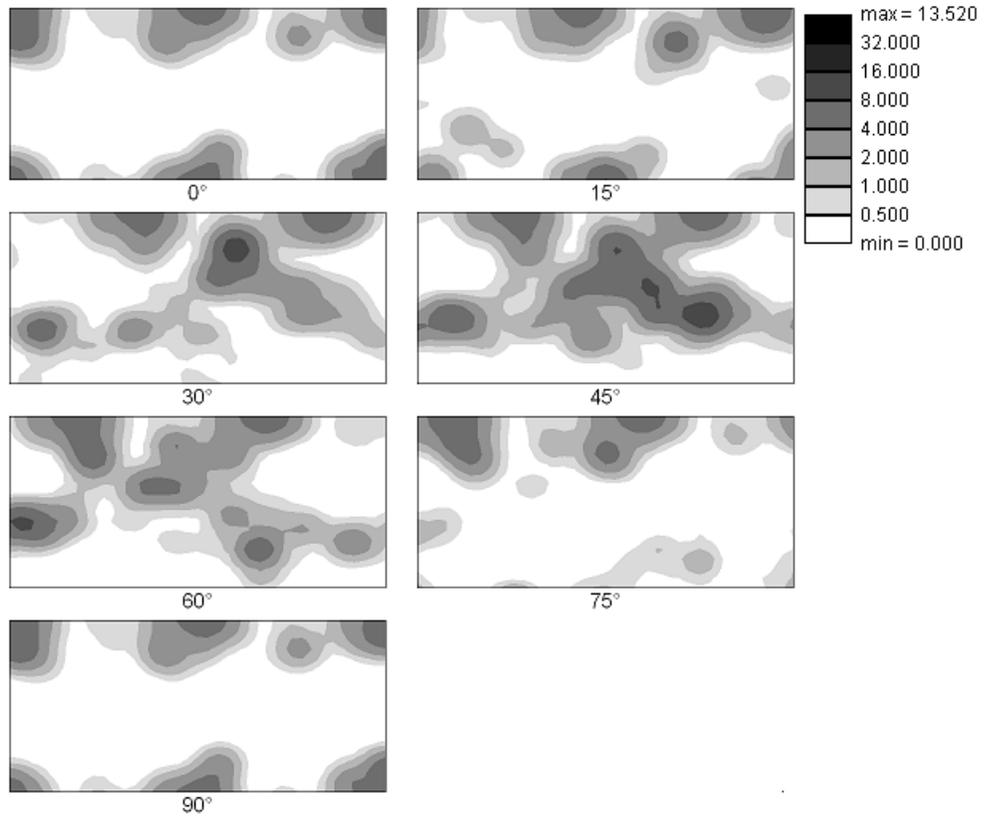

Fig. 8. $\varphi_2$ sections of the ODF of the surface layer of as-received niobium sheets

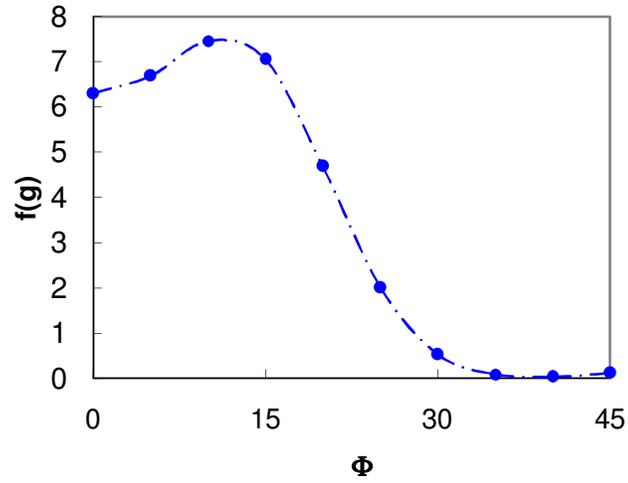

Fig. 9. The skeleton diagram showing the $\eta$ fiber ($\varphi_1 = \varphi_2 = 0$) of surface layer of the as-received niobium.

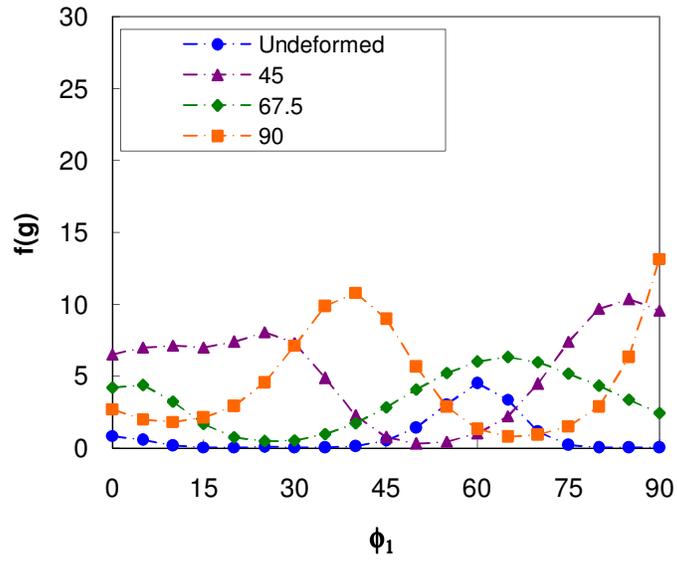

(a)

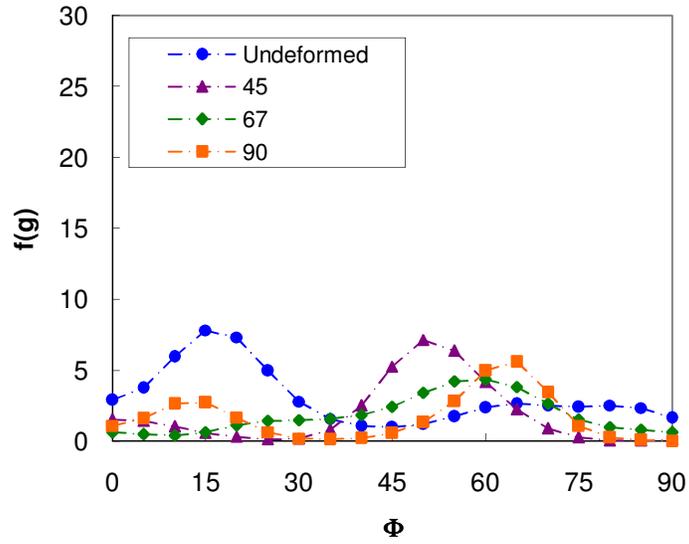

(b)

Fig. 10. The skeleton diagrams showing a) the $\gamma$ fiber and b) the $\alpha$ fiber of surface layer of the undeformed niobium and uniaxially deformed niobium in 45, 67.5, and 90° directions

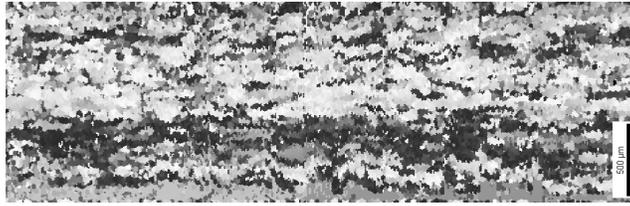

Fig. 11. The through thickness inverse pole figure map of the bulged niobium sample cut in 0°

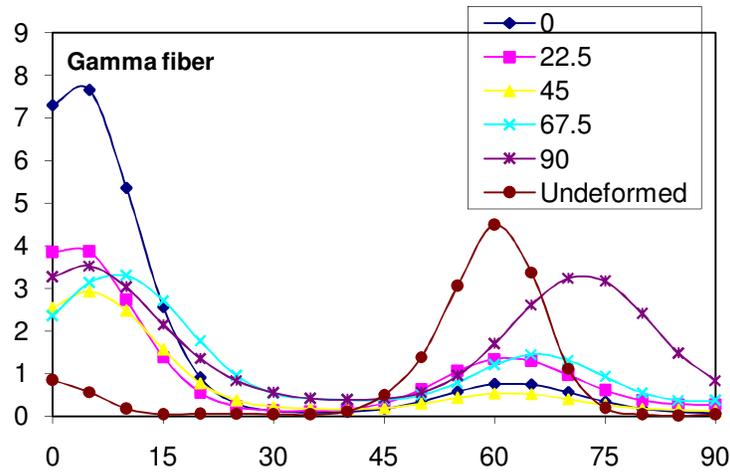

(a)

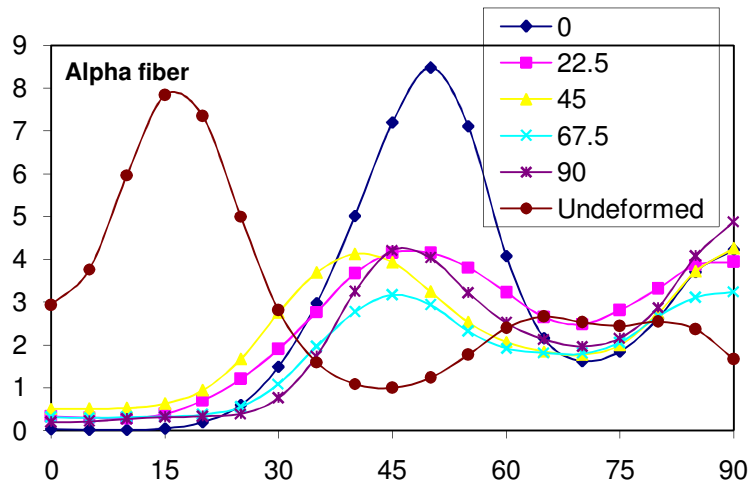

(b)

Fig. 12. The skeleton diagrams showing a) the $\gamma$ fiber and b) the $\alpha$ fiber of surface layer of the bulged niobium in 0°, 22.5°, 45°, 67.5°, and 90° directions

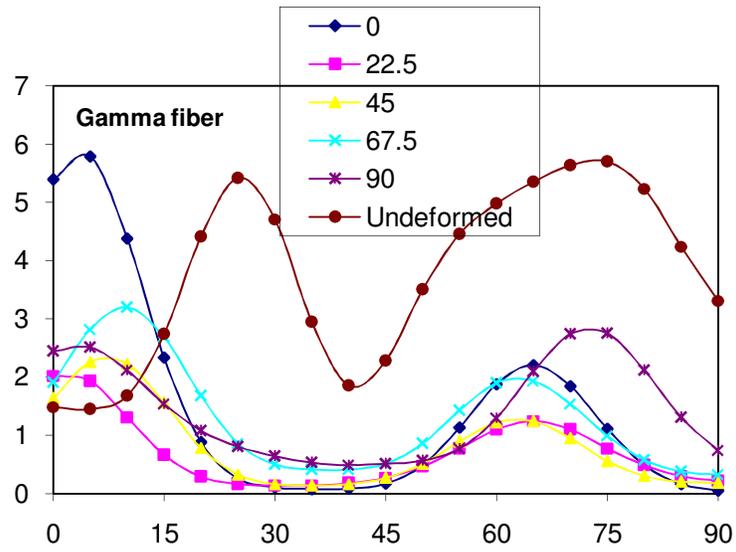

(a)

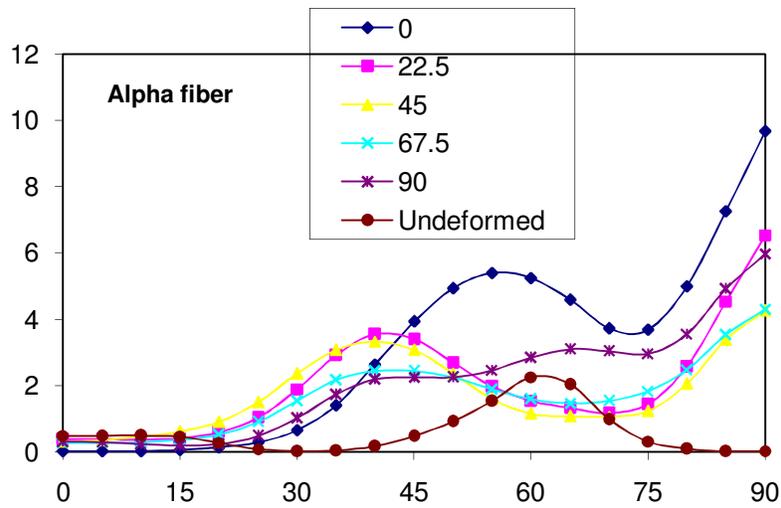

(b)

Fig. 13. a) the $\gamma$ fiber and b) the $\alpha$ fiber of subsurface layer of the bulged niobium in different directions

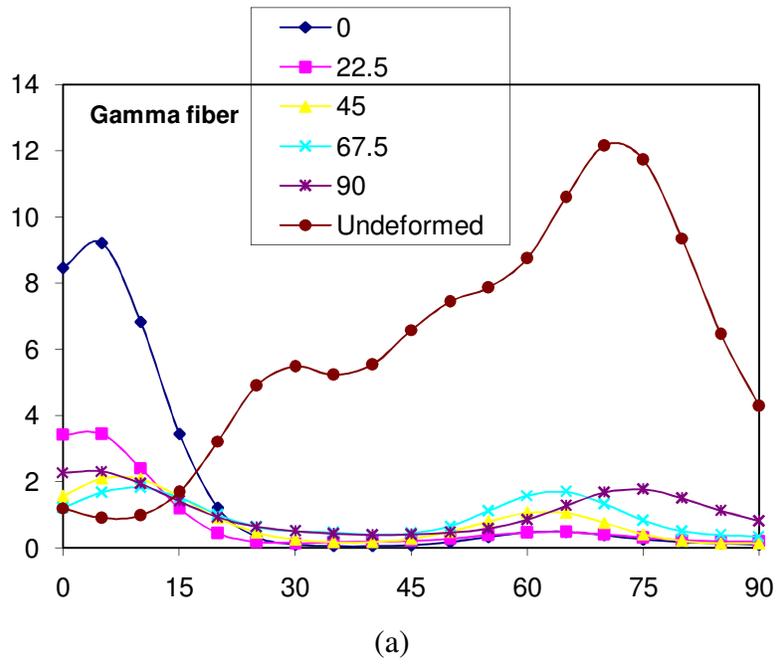

(a)

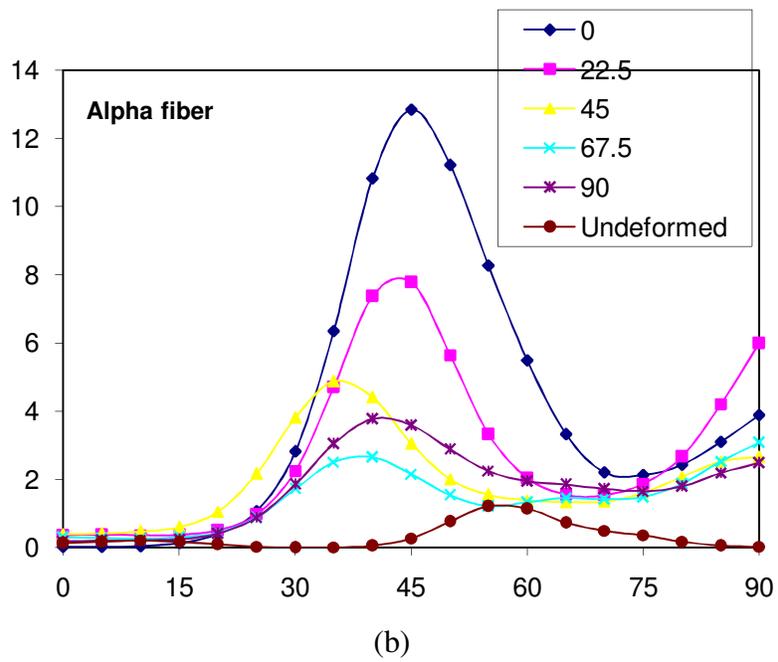

(b)

Fig. 14. a) the $\gamma$ fiber and b) the $\alpha$ fiber of mid-thickness layer of the bulged niobium